\documentclass[twocolumn,showpacs,preprintnumbers,amsmath,amssymb,prb]{revtex4}
\usepackage{graphicx} 
\usepackage{dcolumn} 
\usepackage{bm}
\usepackage{subfigure} 
\usepackage{float} 
\usepackage{color}

\begin{document}

\title{\bf{ Thermodynamics of the one-dimensional frustrated Heisenberg ferromagnet with 
arbitrary spin}}

\author{M. H\"{a}rtel}
\author{J. Richter}
\affiliation{Institut f\"{u}r Theoretische Physik, Otto-von-Guericke-Universit\"{a}t Magdeburg, D-39016 Magdeburg, Germany}
\author{D. Ihle}
\affiliation{Institut f\"{u}r Theoretische Physik, Universit\"{a}t Leipzig, D-04109 Leipzig, Germany}
\author{J. Schnack}
\affiliation{Universit\"at Bielefeld, Fakult\"at f\"ur Physik,
  Postfach 100131, D-33501 Bielefeld, Germany}
\author{S.-L. Drechsler}
\affiliation{Leibniz-Institut f\"{u}r Festk\"{o}rper- und Werkstoffforschung Dresden, D-01171 Dresden, Germany}

\date{\today}

\begin{abstract}
The thermodynamic quantities (spin-spin correlation functions $\langle
{\bf S}_0{\bf S}_n\rangle$,
correlation length $\xi$,
spin  susceptibility $\chi$, and specific heat $C_V$) of the frustrated one-dimensional 
$J_1$-$J_2$ Heisenberg ferromagnet with arbitrary spin quantum number $S$ 
below the quantum critical point, i.e. for $J_2< |J_1|/4$, are calculated 
using a rotation-invariant
Green-function 
formalism and full diagonalization as well as a finite-temperature
Lanczos technique   for finite chains of up to $N=18$ sites. 
The low-temperature behavior of the susceptibility $\chi$ and the correlation length
$\xi$ is well described by 
$  \chi =\frac{2}{3}S^4\left(|J_1|-4J_2\right) T^{-2} + A
S^{5/2}\left(|J_1|-4J_2\right)^{1/2} T^{-3/2} 
$
and
$  \xi =S^2\left(|J_1|-4J_2\right) T^{-1} + B S^{1/2}
\left(|J_1|-4J_2\right)^{1/2} T^{-1/2}
$ with $A \approx 1.1 \ldots 1.2$ and $B \approx 0.84 \ldots 0.89$.
The vanishing of the factors in front of the temperature  at $J_2=|J_1|/4$
indicates  a change of the critical behavior  
of $\chi$ and $\xi$ at $T \to 0$.
The specific heat may exhibit an 
additional  frustration-induced  
low-temperature maximum 
when approaching the quantum critical point. This maximum appears for
$S=1/2$ and $S=1$, but was not found for $S>1$.
\end{abstract}

\maketitle

\section{ Introduction} \label{intro}
The one-dimensional (1D) Heisenberg model with ferromagnetic
nearest-neighbor (NN) exchange $J_1<0$ and frustrating antiferromagnetic
next nearest-neighbor (NN) exchange $J_2 \ge 0$ has recently attracted much
attention, see, e.g., 
Refs.~\onlinecite{bader79,chubukov,honecker,krivnov07,krivnov08,hiki,haertel08,zinke09,lauchli09,sirker2010,haertel11,nishimoto}.
This model may serve as the minimal model to describe the physical
properties of edge-shared chain
cuprates, e.g. Ca$_2$Y$_2$Cu$_5$O$_{10}$, 
Li$_2$ZrCuO$_4$, 
and  Li$_2$CuO$_2$.\cite{Kudo05,drechsQneu,malek,nishimoto}
On the other hand, several materials considered as  1D ferromagnets, 
such as 
TMCuC[(CH$_3$)$_4$NCuCl$_3$],\cite{TMCuC}, CsNiF$_3$,\cite{steiner}
Ni(4-cyanopyridine)$_2$Cl$_2$\cite{landee} or deposited Co
chains\cite{Co_chains} 
might have also a weak frustrating  
NNN exchange $J_2 < -J_1/4$.
 Moreover, recent investigations suggest that
Li$_2$CuO$_2$ as well as  
 Ca$_2$Y$_2$Cu$_5$O$_{10}$
are quasi-1D spin systems with a dominant ferromagnetic $J_1$  and
weak frustrating
antiferromagnetic  $J_2$  so that the inchain spin-spin correlations are predominantly
ferromagnetic.\cite{Kudo05,malek,nishimoto}

The corresponding Hamiltonian of the 1D $J_1$-$J_2$ Heisenberg model considered in this
paper is given by
\begin{equation}  \label{Ham}
  H=\sum_{i} \left(J_1 {\bf S}_i{\bf S}_{i+1}+J_2{\bf S}_i{\bf
S}_{i+2}\right ) ,
\end{equation}
where $i$ runs over all lattice sites and ${\bf S}_i^2=S(S+1)$.
We set $J_1=-1$ and consider $J_2 \ge 0$.  
Although some of the above mentioned materials, namely 
CsNiF$_3$,\cite{steiner}
Ni(4-cyanopyridine)$_2$Cl$_2$\cite{landee} or deposited Co
chains\cite{Co_chains}, represent 1D ferromagnets with spin quantum number
$S > 1/2$,
so far the focus of recent theoretical studies has been on the $S=1/2$ case.
Since Haldane's famous paper\cite{haldane} we know that in 1D Heisenberg
systems the spin
quantum number may play a crucial role. 
Recently it has been found that 
the frustrated model (\ref{Ham}) in the extreme quantum case $S=1/2 $ may exhibit
a different behavior near the quantum critical point than the model for
$S>1/2$.\cite{krivnov07}
Moreover,  in
Ref.~\onlinecite{ihle_new} it has been found
that
for the unfrustrated 1D quantum ferromagnets a  characteristic 
field-induced low-temperature maximum in the specific heat 
exists only  for the small
spin quantum numbers $S=1/2$ and $S=1$, see also
Refs.~\onlinecite{haertel11,magfeld,antsyg}.

In the present paper we discuss the thermodynamics of the model (\ref{Ham})
for arbitrary spin quantum number $S$ and focus on the 
parameter regime $J_2<|J_1|/4$, where the ferromagnetic ground state is
realized. 
It has been recently demonstrated for low-dimensional frustrated
ferromagnets,\cite{haertel08,haertel10,dmitriev_clas,haertel11}
that the frustrating
$J_2$ may influence the thermodynamics  substantially.
In the 1D system
a change in the low-temperature
behavior of the susceptibility and the correlation length
as well as an additional
low-temperature maximum
in the specific heat have been found when approaching the zero-temperature
critical point at $J_2=|J_1|/4$
from the ferromagnetic side.\cite{haertel08}
In particular, a different critical behavior of the susceptibility and the
correlation length has been found for $J_2 < |J_1|/4$ and at $J_2 =
|J_1|/4$.\cite{haertel08,dmitriev_clas}

As in our previous papers on frustrated spin-$1/2$
ferromagnets,\cite{haertel08,haertel10,haertel11} 
we use a second-order
Green-function technique\cite{kondo} to study the influence of the spin
quantum number $S$ on the thermodynamic properties of the model  (\ref{Ham}).
This technique 
has been applied successfully to 
several low-dimensional quantum spin systems.
\cite{rgm_new,magfeld,antsyg,ihle_new,junger05,junger09,haertel08,haertel10,haertel11,kondo,SSI94}

To extend our Green's function theory\cite{haertel08} to $S>1/2$, 
we follow Ref.~\onlinecite{junger09}, 
where the Green's function technique was 
applied to a ferro- and antiferromagnetic layered square 
lattice with arbitrary $S$.
We complement the Green's function results by 
full exact diagonalization (ED) and finite-temperature Lanczos (FTL) technique data
for finite systems of up 
to $N=18$ lattice sites.

\section{Full diagonalization and finite-temperature Lanczos technique for finite lattices} 
Using Schulenburg's 
{\it spinpack}\cite{spinpack} and exploiting
the lattice symmetries and the fact that  $S^z=\sum_i S_i^z$ 
commutes with $H$
we are able to calculate
the exact thermodynamics for periodic chains  of up to $N=14$ spins for spin
quantum number $S=1$. For $N=8$ sites we found the exact thermodynamics up
to $S=2$. 
Clearly, the full ED of finite systems for $S>1/2$ is 
less efficient as for $S=1/2$, since the accessible system size $N$
decreases with increasing of $S$.

In addition to the full ED we have  also applied a
FTL technique, see e.g.
Refs.~\onlinecite{prelov,schnack}. This method allows an accurate calculation
of thermodynamic quantities down to low temperatures for $S=1$ up to $N=18$.

\section{Spin-rotation-invariant Green-function theory} \label{RGM}
To study the  thermodynamics of the model (\ref{Ham}) for arbitrary $N$ we use the spin-rotation-invariant
Green-function method (RGM). 
\cite{haertel08,kondo,rgm_new,haertel10,junger09,SSI94} The relevant
thermodynamic quantities can be determined by calculating the two-time commutator 
Green function \cite{elk} $\langle\langle
S_q^+;S_{-q}^-\rangle\rangle_\omega$, 
which is related to
the transverse 
spin susceptibility by  
$\chi_q^{+-}\left(\omega\right)=
- \langle\langle S_q^+;S_{-q}^-\rangle\rangle_\omega$ .
Using the equations of motion up to the second step and supposing rotational symmetry
with $\langle
S_i^z\rangle=0$,  we obtain
$
  \omega^2\langle\langle S_q^+;S_{-q}^-\rangle\rangle_\omega=M_q+\langle\langle -\ddot{S}_q^+;S_{-q}^-\rangle\rangle_\omega
$
with $M_q=\langle\left[\left[S_q^+,H\right],S_{-q}^-\right]\rangle$ 
and $-\ddot{S}_q^+=\left[\left[S_q^+,H\right],H\right]$. 
For the moment $M_q$  the exact expression
\begin{equation}\label{moment}
  M_q=-4\sum_{n=1,2}J_nC_n\left(1-\cos nq\right)
\end{equation}
holds,
where $C_n=\langle S_0^+S_n^-\rangle=2\langle S_0^zS_n^z\rangle$. The second derivative $-\ddot{S}_q^+$ 
is approximated in the spirit of 
Refs.~\onlinecite{SSI94,kondo,rgm_new,magfeld,antsyg,haertel10,ihle_new,junger05,junger09},
i.e.,
in $-\ddot{S}_i^+$ we use the decoupling $S_i^+S_j^+S_k^- = 
\alpha\langle S_j^+S_k^-\rangle S_i^++\alpha\langle S_i^+S_k^-\rangle S_j^+$,
and $S_i^+S_j^-S_j^+ = 
\langle S_j^-S_j^+\rangle S_i^+ + \lambda\langle S_i^+S_j^-\rangle S_j^+$ for products with two 
coinciding sites which occur for $S\geq
1$.\cite{ihle_new,junger05,junger09,SSI94}
Note that for $J_2<-J_1/4$, 
where the ground-state  is ferromagnetic,
the vertex parameters $\alpha$ and $\lambda$ can be assumed in a good approximation 
to be independent 
of the range of the associated spin
correlators, cf. Ref.~\onlinecite{haertel08}.  Then we obtain $-\ddot{S}_q^+=\omega_q^2S_q^+$ and
\begin{equation}\label{Greenfunction}
  \chi_q^{+-}\left(\omega\right)= -\langle\langle
S_q^+;S_{-q}^-\rangle\rangle_\omega=\frac{M_q}{\omega_q^2-\omega^2} ,
\end{equation}
with
\begin{equation}\label{dispersion}
\omega_q^2= \hskip-3mm \sum_{n,m(=1,2)} \hskip-3mm
J_nJ_m\left(1-\cos nq\right)
\left[K_{n,m}+4\alpha C_n\left(1-\cos mq\right)\right],
\end{equation}
where $K_{n,n}=\frac{4}{3}S\left(S+1\right)+2\lambda C_n+2\alpha\left(C_{2n}-3C_n\right)$,
$K_{1,2}=2\alpha\left(C_3-C_1\right)$,
and $K_{2,1}=K_{1,2}+4\alpha\left(C_1-C_2\right)$.
From the Green function (\ref{Greenfunction}) the correlation functions $C_n=\frac{1}{N}
\sum_qC_qe^{iqn}$ are determined by 
the spectral theorem,\cite{elk}
\begin{equation} \label{spectral}
  C_q=\langle
S_q^+S_{-q}^-\rangle=\frac{M_q}{2\omega_q}\left[1+2n
\left(\omega_q\right)\right],
\end{equation}
where $n(\omega_q)=\left(e^{\omega_q/T}-1\right)^{-1}$ is the Bose function. 
Using the operator identity ${\bf S}_i^2=S_i^+S_i^--S_i^z+\left(S_i^z\right)^2$, we get the 
sum rule $C_0=\frac{1}{N}\sum_qC_q=\frac{2}{3}S\left(S+1\right)$. 
Following Ref.~\onlinecite{junger09}, as an additional equation to determine the vertex parameters
 we  consider the ratio
\begin{equation}\label{rellambda}
  r(T)=\frac{\lambda(T)-\lambda(\infty)}{\alpha(T)-\alpha(\infty)}=r(0)
\end{equation}
as temperature independent, where $\lambda(\infty)=1-\frac{3}{4S(S+1)}$ and
$\alpha(\infty)=1$, see Ref.~\onlinecite{junger05}.
The uniform static 
spin susceptibility $\chi=\lim_{q\to 0}\chi_q$, where 
$\chi_q=\chi_q\left(\omega=0\right)$ and $\chi_q\left(\omega\right)=\frac{1}{2}\chi_q^{+-}\left(\omega\right)$, 
is given by
\begin{equation}\label{susceptibility}
  \chi=-\frac{2}{\Delta}\sum_{n=1,2}n^2J_nC_n\,; \quad \Delta=\hskip-1mm
\sum_{n,m(=1,2)}\hskip-1mm 
n^2 J_nJ_mK_{n,m}.
\end{equation}
The correlation length $\xi$ for a ferromagnet can be calculated from the expansion 
of the static spin susceptibility around $q=0$ 
(see, e.g., Refs.~\onlinecite{kondo} and \onlinecite{ihle_new}), $\chi_q=\chi/\left(1+\xi^2q^2\right)$.
The ferromagnetic long-range order, occurring in the 1D model at zero
temperature  only, is related to the condensation 
term $C$ (see Ref.~\onlinecite{kondo}) via  $C_n\left(0\right)=\frac{1}{N}\sum_{q(\neq 0)}\left(M_q/2\omega_q\right)e^{iqn}+C$. 
Equating this expression  to the 
exact result 
$C_{n}\left(0\right)=\frac{2}{3}S\delta_{n,0}+\frac{2}{3}S^2$ leads to
$C(0)=\frac{2}{3}S^2$ and
$M_q(0)/2\omega_q(0)=\frac{2}{3}S$.\cite{junger09}
This requires $\alpha(0)=\frac{3}{2}$ and $K_{n,m}(0)=0$ [cf. Eqs.~(\ref{moment}) and
(\ref{dispersion})] which leads to $\lambda(0)=2-\frac{1}{S}$.
The parameter $\Delta$ in Eq.~(\ref{susceptibility}) is zero at $T=0$,
i.e., $\chi$ diverges as $T\to 0$ 
according to the ferromagnetic phase transition.
Moreover, using these results we find at zero temperature
 $\omega_{q}=2\rho_s{q}^2$ ($|{q}| \ll 1$), $\rho_s
=\frac{S}{2}(|J_1|-4J_2)$, where $\rho_s$ is the
spin stiffness.

Since we will compare the RGM data with ED data for finite lattices,
see Sec.~\ref{results},  we have to adopt the RGM to finite
$N$.  In this case the quantity $C$ is not related to the 
magnetisation and stays nonzero in the whole temperature region. 
In Ref.~\onlinecite{junger05} it was shown that $C=2T\chi/N$.

Finally, to solve the equation of motion and to evaluate the thermodynamic quantities
we have to determine the correlators $C_l$
($l=1,\ldots,4$) and the vertex parameters $\alpha$ and $\lambda$ (for finite
systems, also $C$) 
as the numerical solution of  
a coupled system of six (seven)  
non-linear algebraic self-consistency equations: $C_l$ according to
Eq.~(\ref{spectral}) 
including the sum rule $C_0=\frac{1}{2}$, and the ratio
Eq.~(\ref{rellambda}).
To find the numerical solution of the RGM equations for $T> 0$,  we start
at high temperatures and decrease $T$ in small steps. Below a certain
(low) temperature $T_0(J_2,S)$ no solutions of the RGM equations (except at
$T=0$) could be found,  since the
quantity $\Delta(T,J_2,S)$ in Eq.~(\ref{susceptibility})
becomes exponentially small
which leads to numerical uncertainties. We find that $T_0(J_2,S)/S(S+1)$ is of the
order of $10^{-3}$ for all values of $J_2$ and $S$ considered here.
Moreover, we find that the RGM solution at very low temperatures
becomes less trustworthy for $J_2$ approaching the quantum critical point 
$J_2 = |J_1|/4$, cf. also
Refs.~\onlinecite{haertel08} and \onlinecite{haertel11}. Therefore,
below we will present  
RGM results for $J_2 \le 0.22|J_1|$ only.
\begin{figure}[ht]
  \begin{center}
    \includegraphics[height=17cm]{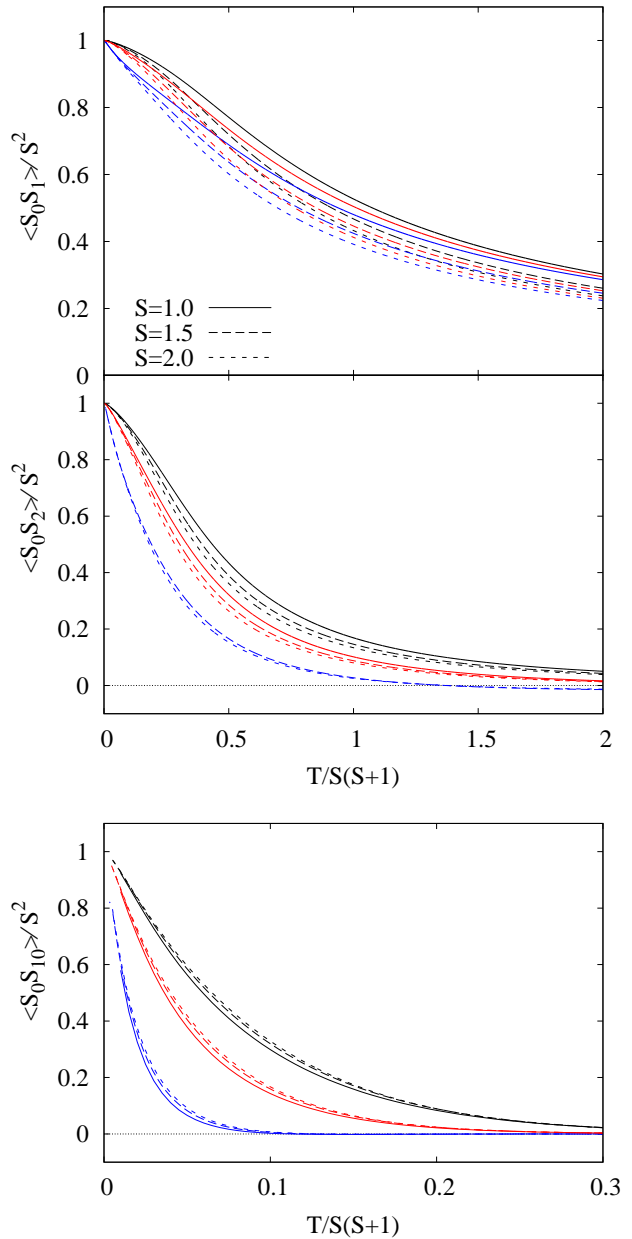}
  \caption{Spin-spin correlation functions 
$\langle {\bf S}_0{\bf S}_{1}\rangle/S^2$ (NN), 
$\langle {\bf S}_0{\bf S}_{2}\rangle/S^2$ (NNN), and
$\langle {\bf S}_0{\bf S}_{10}\rangle/S^2$  obtained by the RGM
for spin 
quantum numbers $S=1$ (solid), $3/2$ (long-dashed), and $S=2$ (short-dashed) 
and  frustration parameters $J_2=0$ (black), $0.1$ (red), and $0.2$
(blue).}\label{fig_korr}
  \end{center}
\end{figure}

\section{Results}\label{results}
Motivated by the high-temperature behavior of the physical
quantities\cite{domb_green,OHZ06} we use
as the renormalized temperature scale $t=T/S(S+1)$, since for large temperatures physical
quantities exhibit $t$-dependence  which is independent of $S$.

\subsection{Spin-spin correlation functions} 
Let us first consider the spin-spin correlation functions
$\langle{\bf S}_0{\bf S}_{n}\rangle$  depicted in Fig.~\ref{fig_korr} for
NN, NNN
and tenth-nearest neighbors. 
Analogously to the frustrated $S=1/2$ 
chain,\cite{haertel08} with increasing frustration the correlation functions decrease more
rapidly. Obviously the spin quantum number $S$ has only a small influence on
the correlation functions as  functions of $t$. Interestingly, an increase of $S$ yields a slight weakening of the
short-range spin-spin correlation at fixed $t>0$ 
(see upper and middle panels in Fig.~\ref{fig_korr}), but an enhancement of larger-distant
correlations (see lower panel in Fig.~\ref{fig_korr}).
Moreover, the NNN correlation function for $J_2=0.2$ becomes negative 
at  $t =t_0 \approx 1.35, 1.37, 1.38 $ for $S=1,3/2,2$, respectively.
(Note that for $S=1/2$ the corresponding temperature is $t_0 \approx
1.27$.\cite{haertel08})
\begin{figure}[ht]
  \begin{center}
    \includegraphics[height=6cm]{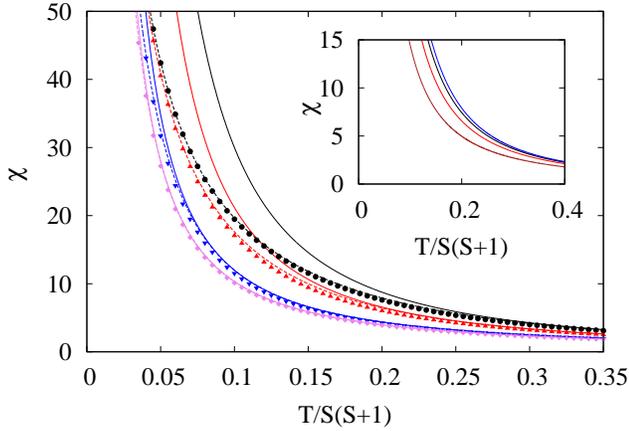}
  \caption{Susceptibility $\chi$ for  spin quantum number $S=1$
and frustration parameters 
$J_2=0$ (black), $0.1$ (red), $0.2$ (blue), and $0.22$ (violet) [solid lines: RGM for
$N\to \infty$, dashed lines: RGM for $N=12$, symbols: ED for $N=12$].
Inset: Susceptibility $\chi$ for $J_2=0.1$  and $S=1/2$, $1$, $3/2$, and $2$ (from bottom to top). 
The data for $S=1/2$ are taken from
Ref.~\onlinecite{haertel08}. 
\label{fig_sus}}
  \end{center}
\end{figure}
\begin{figure}[ht]
  \begin{center}
    \includegraphics[height=6cm]{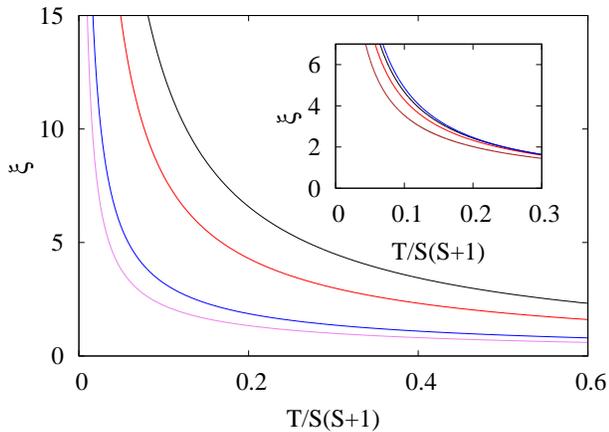}
  \caption{Correlation length $\xi$  
for  spin quantum number $S=1$
and frustration parameters 
$J_2=0$ (black), $0.1$ (red), $0.2$ (blue), and $0.22$ (violet).
Inset: Correlation length $\xi$ 
for $J_2=0.1$  and 
$S=1/2$, $1$, $3/2$, and $2$ 
(from bottom to top). The data for $S=1/2$ are taken from
Ref.~\onlinecite{haertel08}. 
\label{fig_xi}}
  \end{center}
\end{figure}

\subsection{Susceptibility and correlation length}
The behavior of the correlation functions 
is reflected in
the susceptibility 
$\chi$ and the correlation length $\xi$ shown in Figs.~\ref{fig_sus} and
\ref{fig_xi}. 
We illustrate 
the influence of frustration for a fixed
spin quantum number $S=1$, whereas in the insets we show the influence of
$S$ for a fixed frustration parameter
$J_2=0.1$.  
Since the ground state is ferromagnetic, 
both quantities diverge at $T=0$.
With increasing of frustration the rapid increase in $\chi$ and $\xi$  is
shifted to lower temperatures.
The comparison of RGM and ED results for finite chains of $N=12$ presented in
Fig.~\ref{fig_sus} demonstrates a very good agreement of the susceptibility
data obtained by  both methods.
It can also be seen that the
renormalized temperature $t$, where finite-size effects become relevant, 
is shifted to lower values of $t$ with increasing frustration (in
Fig.~\ref{fig_sus} the curves for $N=12$ and $N \to \infty$ for $J_2=0.2$ and
$J_2=0.22$ almost coincide). 
The curves shown in the insets again illustrate that the influence of the
spin quantum number is small; only for $S=1/2$ the curves are noticeably separated
from the bundle of curves for $S>1/2$.
Moreover, it is evident that $\chi$ and $\xi$ at fixed $t$ and $J_2$
increase with growing $S$.
\begin{figure}[ht]
  \begin{center}
    \includegraphics[height=12cm]{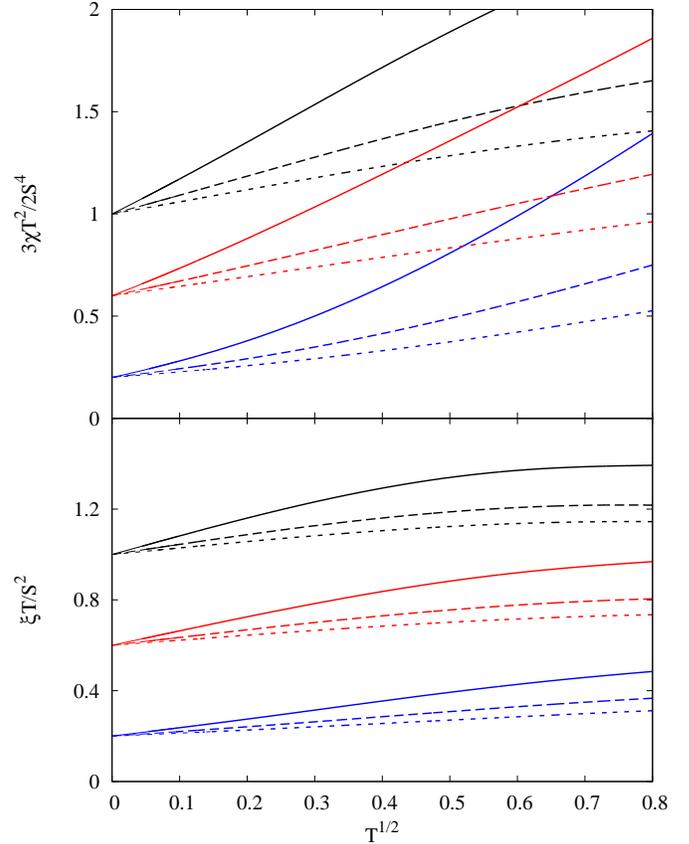}
  \caption{RGM results for $\frac{3}{2}\chi T^2/S^4$ (upper panel) and  
$\xi T/S^2$  (lower panel) versus $\sqrt{T}$  for spin quantum numbers 
$S=1$ (solid), $3/2$ (long-dashed), and $S=2$ (short-dashed) and frustration
parameters $J_2=0$ (black), $0.1$ (red), and $0.2$ (blue).
}
\label{fig_chiTT}
  \end{center}
\end{figure}

Next we consider the critical behavior of $\chi$ and $\xi$ for $T\to 0$ by
analyzing the RGM data at low temperatures.  
From Ref.~\onlinecite{SSI94} it is known 
that within the RGM the low-temperature behavior of $\chi$ and $\xi$ 
of the unfrustrated ferromagnet ($J_2=0$) is given by 
$\lim_{T \to 0}\chi T^2=2S^4|J_1|/3$ and $\lim_{T \to 0}\xi T=S^2|J_1|$.  These results
agree with those obtained by modified spin-wave
theory.\cite{takahashi,yamada2} Moreover, for $S=1/2$ they coincide with the exact
Bethe-ansatz analysis.\cite{yamada1,yamada2} (Note that $\chi$ 
defined  in Refs.~\onlinecite{takahashi}
and
\onlinecite{yamada1}
is larger by a factor of four than $\chi$ given by
Eq.~(\ref{susceptibility}).) 
Including higher-order terms in $T$ for 1D Heisenberg ferromagnets the 
low-temperature behavior of the susceptibility and the correlation length
reads
$  \chi T^2=y_0+y_1\sqrt{T}+y_2T+{\cal O}(T^{3/2})$ and 
$  \xi T=x_0+x_1\sqrt{T}+x_2T+{\cal O}(T^{3/2})$, see, e.g.,
Refs.~\onlinecite{yamada1,yamada2,takahashi,haertel08}. The dependence of $\chi T^2$
and $\xi T$ on $T^{1/2}$ is shown in Fig.~\ref{fig_chiTT}
for various values of $S$ and $J_2$. The validity of linear relations 
$\chi T^2 \propto T^{1/2} $ and $\xi T \propto T^{1/2} $ at low temperatures
is clearly seen. 
To determine the parameters $y_0$, $y_1$, $y_2$, $x_0$, $x_1$, and $x_2$,
we follow the lines of Ref.~\onlinecite{haertel08} and fit the RGM results
for $\chi$ and $\xi$ using the relations given above.
For the fits we use low-temperature data points 
between $T_0$ and  $T_0+T_{cut}$, where $T_0$ is the lowest  temperature
for which solutions of the RGM equations can be found, see Sec.~\ref{RGM}. 
For $T_{cut}$ 
we chose $T_{cut}=0.005$, cf.
Ref.~\onlinecite{haertel08}. 
By analyzing data for $J_2=0$ and $S=1$, $3/2$, $2$ and $5$ we have numerically
confirmed the relations $y_0=2S^4/3$ and $x_0=S^2$ with an accuracy of at least three digits.
For the next-to-leading coefficients we adopt the relations
$y_1/S^4=\alpha_0S^{\alpha_1}$ and $x_1/S^2=\beta_0S^{\beta_1}$ found by
modified spin-wave
theory for the unfrustrated model.\cite{takahashi,yamada2}
We find
$\alpha_0=1.106$, $\alpha_1=-1.497$,  $\beta_0=0.834$, and $\beta_1=-1.495$,
which is  in good  agreement 
with the results of Refs.~\onlinecite{takahashi} and \onlinecite{yamada2},  where
$\alpha_0=0.824$, $\alpha_1=-1.5$, 
$\beta_0=0.412$, and $\beta_1=-1.5$ were reported.

\begin{figure}[ht]
  \begin{center}
    \includegraphics[height=12cm]{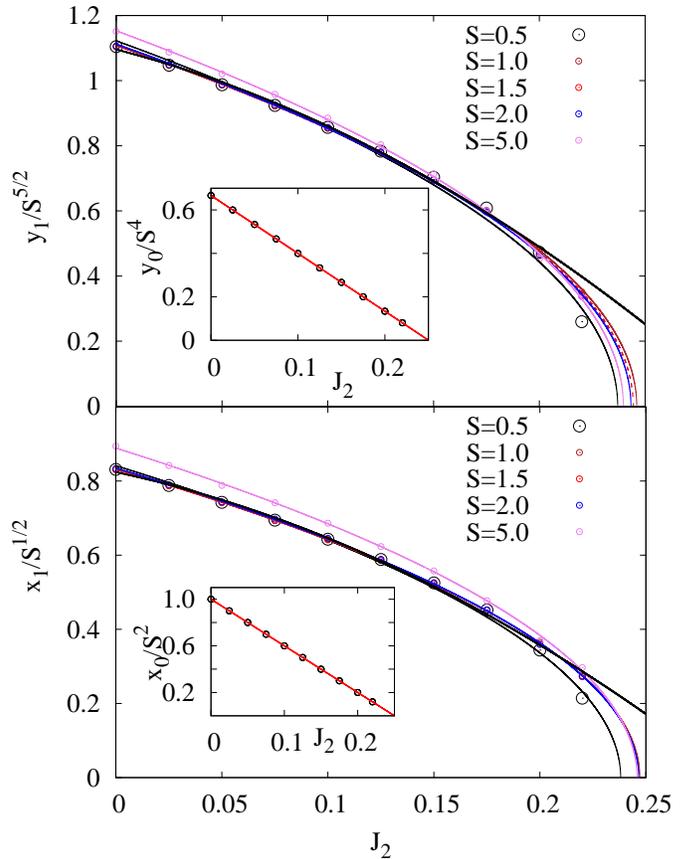}
  \caption{Next-to-leading coefficients
$y_1/S^{5/2}$ (upper panel) and $x_1/S^{1/2}$ (lower panel) in dependence on $J_2$ obtained by fitting
the low-temperature data of $\chi$ and $\xi$ (see main text). The symbols
represent the
data points for $S=1/2$, $1$, $3/2$, $2$, and $S=5$, and the lines  of
the same color show the corresponding fit curves of these points using the fit function
$f(J_2)=a\sqrt{1-bJ_2}$.  The thick black
solid line shows the quadratic fit for $S=1/2$ without the data point
at $J_2=0.22$ as used in Ref.~\onlinecite{haertel08}.\\
Insets: Leading coefficients
$y_0$ (upper panel) and $x_0$ (lower panel) in dependence on $J_2$ obtained by fitting
the low-temperature data of $\chi$ and $\xi$ (see main text). The values for $y_0/S^4$
as well as for  $x_0/S^2$
practically coincide for $S=1/2$, $1$, $3/2$, $2$, and $S=5$ used to
determine $y_0/S^4$ and $x_0/S^2$. 
}
\label{fig_y1_x1}  
\end{center}
\end{figure}

Now we determine the 
parameters $y_0$, $y_1$, $y_2$, $x_0$, $x_1$, and $x_2$
for finite frustration $J_2>0$.
The results for the leading coefficients $y_0$ and $x_0$ 
are shown in the insets of 
Fig.~\ref{fig_y1_x1}. 
Both coefficients $y_0$ and $x_0$ obey with high precision  the linear relations 
$y_0/S^4=\frac{2}{3}\left(|J_1|-4J_2\right)\,$  and
$x_0/S^2=|J_1|-4J_2$, 
i.e., the leading coefficients decrease with  growing frustration
and vanish finally at the quantum critical point
$J^c_2=|J_1|/4$, thus
indicating a change of the critical exponent of $\chi$ and $\xi$ at
$J^c_2$.  
This linear decrease  of 
$y_0$ and $x_0$
found by fitting the low-temperature behavior of $\chi$ and $\xi$
is the same as that obtained analytically for the zero-temperature
spin stiffness
$\rho_s$, see
Sec.~\ref{RGM}.  This correspondence between the spin stiffness
and the divergence
of  the susceptibility  and the correlation length is in accordance with
general arguments\cite{ivanov,kopietz91}
concerning the low-temperature physics of low-dimensional Heisenberg
ferromagnets.
Moreover, these linear relations  agree with the findings for $S=1/2$
(Ref.~\onlinecite{haertel08}) and $S \to \infty$
(Ref.~\onlinecite{dmitriev_clas}).

Next we determine the coefficients 
$y_1$ and $x_1$ as functions of $J_2$
again using 
the  fit of the RGM data for $S=1$, $3/2$, $2$ and $5$ described above.
Moreover, we reanalyze our
data for $S=1/2$ from Ref.~\onlinecite{haertel08}. Using an improved
computer code we are able to add  $S=1/2$ data for $J_2=0.22$ not given in
Ref.~\onlinecite{haertel08}.
Similar as for $y_0/S^4$ and $x_0/S^2$, the data for $y_1$ and $x_1$ shown in
Fig.~\ref{fig_y1_x1} yield evidence for  
a universal $J_2$ dependence  of $y_1/S^{5/2}$ and $x_1/S^{1/2}$. 
However, the coincidence of the data points is less pronounced (in
particular for $S=5$) than for $y_0$ and $x_0$.
The dependence  of $y_1/S^{5/2}$ and $x_1/S^{1/2}$ on $J_2$ is
clearly not linear.
As suggested  by the behavior of $y_1/S^{5/2}$ and $x_1/S^{1/2}$ at larger $J_2$ we chose
$f(J_2)=a\sqrt{1-bJ_2}$ to fit the data points presented in 
Fig.~\ref{fig_y1_x1}. These fits suggest, that 
$y_1$ and $x_1$ also vanish approaching the quantum critical point. Indeed, we
find for the fit
parameter $b\approx 4.08$ for most of the fits; only  for $S=1/2$
we have $b\approx 4.20$.      
For comparison, we also show the previous quadratic fit
for $S=1/2$ of Ref.~\onlinecite{haertel08}, where the data point
at $J_2=0.22$ was not included. It is obvious that the square-root fit
is more reasonable than the qadratic fit 
(which leads to a finite $y_1$ and $x_1$ at $J_2=|J_1|/4$). 
Moreover, a square-root dependence of $y_1$ and
$x_1$ on the exchange couplings is also suggested by modified spin-wave
theory.\cite{takahashi,yamada2,sirker2011} 
Based on the  results for 
 $y_0$, $y_1$, $x_0$,  and $x_1$ discussed above we finally argue that 
the low-temperature behavior of the susceptibility  and the correlation length
for $J_2 < |J_1|/4$ is well described by 
\begin{equation} 
  \chi T^2=\frac{2}{3}S^4\left(|J_1|-4J_2\right)+ A
S^{5/2}\sqrt{\left(|J_1|-4J_2\right)}\sqrt{T} 
\end{equation} 
and 
\begin{equation}
  \xi T=S^2\left(|J_1|-4J_2\right)+ B S^{1/2} \sqrt{\left(|J_1|-4J_2\right)}\sqrt{T}
\end{equation}
with $A \approx 1.1$ ($A \approx 1.2$) and $B \approx 0.84$ ($B \approx 0.89$) for $S=1/2$, $1$, $3/2$, and $2$
($S=5$).
Therefore, we conclude that, in accordance with  previous results for $S=1/2$ and 
$S = \infty$, see Refs.~\onlinecite{haertel08} and
\onlinecite{dmitriev_clas}, the critical
behavior of $\chi$ and $\xi$ is changed at the quantum critical point over the
entire range of spin quantum numbers, $1/2 \le S \le \infty$.
We may speculate, that the expansion $  \chi T^2=y_0+y_1\sqrt{T}+y_2T+{\cal
O}(T^{3/2})$ and
$  \xi T=x_0+x_1\sqrt{T}+x_2T+{\cal O}(T^{3/2})$ valid for $J_2 < | J_1|/4$
breaks down at the quantum critical point $J^c_2=|J_1|/4$, and the  critical behavior found for  
the classical system   at $J^c_2$, 
$\chi\propto T^{-4/3}$ and  $\xi\propto T^{-1/3}$, may hold also for the
quantum model. However, recent numerical studies\cite{sirker2011} indicate a
slightly different critical behavior for $S=1/2$.

\subsection{Specific heat}
\begin{figure}[ht]
  \begin{center}
    \includegraphics[height=5cm]{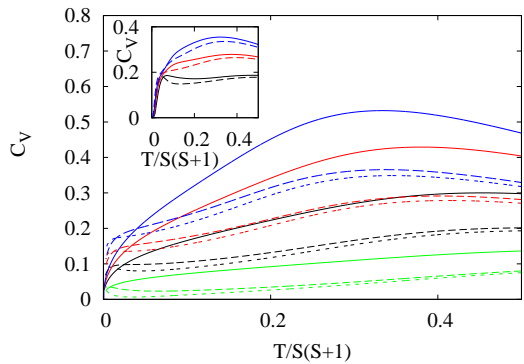}
  \caption{Specific heat for $S=1/2$ (green), 
$S=1$ (black), $S=3/2$ (red), 
and $S=2$ (blue) obtained by RGM for $J_2=0$ (solid), $J_2=0.2$ (long-dashed), and $J_2=0.22$
(short-dashed). 
Inset: ED results for $N=8$ and $S=1$ (black), $3/2$ (red) and
$2$ (blue) for $J_2=0.2$ (solid) and $J_2=0.22$ (dashed).} \label{fig_C}
  \end{center}
\end{figure}

\begin{figure}[ht]
  \begin{center}
    \includegraphics[height=10cm]{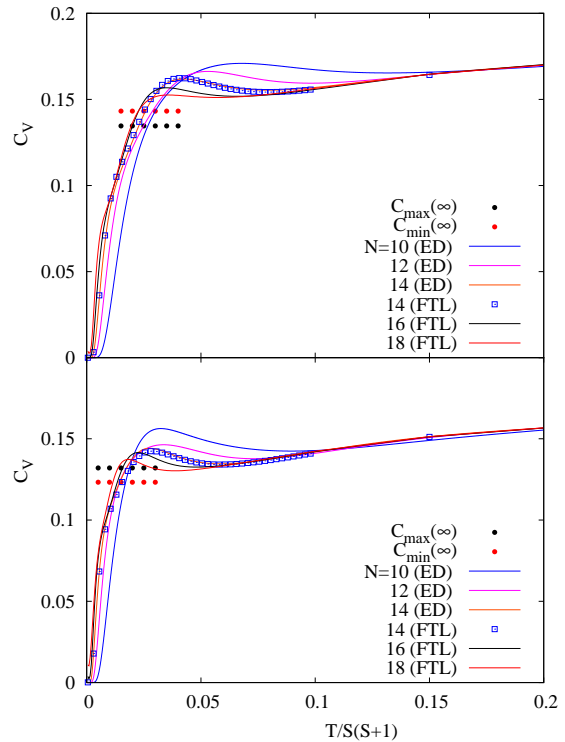}
  \caption{Exact diagoalization (ED)  and 
finite-temperature Lanczos (FTL) results for the low-temperature maximum in the specific heat for $S=1$ and
$J_2=0.2$ (upper panel) and $J_2=0.22$ (lower panel) for chain lengths
$N=10,12,14,16$ and $18$.
For $N=14$ we present both the exact ED and the approximate FTL data to demonstrate
the accruracy of the FTL technique down to very low temperatures.  
}
\label{fig_C_ED}
  \end{center}
\end{figure}

Finally we discuss the specific heat $C_V$, see Figs.~\ref{fig_C} and
\ref{fig_C_ED}.
With increasing of $S$ there is a shift of the
broad maximum typical for spin systems
to lower values of the renormalized temperature $t=T/S(S+1)$ and an increase of its height, see Fig.~\ref{fig_C}.
For fixed $S$ the frustration $J_2$ leads to a decrease of this maximum.
Of particular interest is the low-temperature behavior of the specific heat.
In Ref.~\onlinecite{haertel08}  
an additional frustration-induced  low-temperature maximum in the
specific heat 
 was found for the $S=1/2$ model studied by RGM and ED.
From Fig.~\ref{fig_C} it is obvious that for higher values of $S$ this
additional maximum disappears. Note that this observation is similar to the findings in
Ref.~\onlinecite{ihle_new} for a field-induced low-temperature maximum in $C_V$,
which appears for low values of $S$ only. 
To be more specific, we find that for $S>1$ no extra maximum appears in the
RGM data for $C_V$ in the whole range of $0<J_2\le 0.22$ accessible within the RGM
approach.
The frustration-induced  low-temperature maximum for $S=1/2$
appears in the RGM for  $J_2\geq 0.16$
(Ref.~\onlinecite{haertel08}). For $S=1$ we find that  a double-maximum
structure in $C_V$ calculated by RGM appears  for $J_2 \geq 0.21$.

These RGM based results are supported by finite-size data (see inset of
Fig.~\ref{fig_C}), i.e. we find also in the ED data for $N=8$ a clear tendency to suppress the extra
low-temperature maximum if $S$ increases.
In more detail we analyze finite-size data for $S=1$, where 
the larger systems are accessible by numerical
methods.
In Fig.~\ref{fig_C_ED} we illustrate the finite size-effects in $C_V$ for
$S=1$ and $J_2=0.2$ and $0.22$. It is obvious that the extra low-temperature
maximum 
is appreciably
affected by finite-size effects.
However, from Fig.~\ref{fig_C_ED}
it is also evident that the height of the extra low-temperature maximum as
well as the nearby minimum behave
monotonously with $N$.
Hence a finite-size extrapolation of the height $C_{max}(N)$ 
of the extra maximum and the related minimum  $C_{min}(N)$ is reasonable.
Analogously to Ref.~\onlinecite{haertel08} we find that
$a(N)=a_0+a_1/N^2 + a_2/N^4$ as a reasonable extrapolation scheme.
The results of the extrapolation are shown in Fig.~\ref{fig_C_ED} by black
[for the extrapolated value $C_{max}(\infty)$] and red [for the extrapolated value
of $C_{min}(\infty)$] filled circles. These extrapolated data for $C_{max}$ and
$C_{min}$ indicate, that there is most likely no extra
low-temperature maximum for $J_2=0.2$, but such a maximum appears for
$J_2=0.22$. Hence, the finite-size data support  
the RGM predictions that the
double-peak structure appears for $S=1$ at $J_2 \geq 0.21$.

\section{Summary} 
In this paper we have studied the influence of a frustrating NNN coupling
$J_2$ on
the thermodynamics of the 1D spin-$S$ $J_1$-$J_2$ Heisenberg model with
ferromagnetic $J_1$ and antiferromagnetic $J_2$ with $J_2 \le |J_1|/4$.
For that we have used
the RGM for infinite chains and  full ED
as well as the FTL
technique  for finite chains. 
We find a universal dependence of the thermodynamic quantities on the renormalized temperature
$t=T/S(S+1)$ at large temperatures. At low temperatures  such a universal behavior is found for the
critical properties of the susceptibility  $\chi T^2/S^4 =(2/3) (|J_1|-4J_2)$ 
and
the correlation length $\xi T/S^2 =|J_1|-4J_2$, i.e.,      
the critical exponents of $\chi$ and $\xi$ for $T \to 0$ are not changed by a frustrating
$J_2 < 0.25|J_1|$. However, our data suggest that at the quantum critical
point $J_2=|J_1|/4$ the critical behavior of $\chi$ and $\xi$ is changed.

For $S=1/2$ and $S=1$ an
additional low-temperature maximum in the specific heat $C_V$  emerges when $J_2$
approaches the quantum critical point. Since we did not observe such an
additional maximum in $C_V$ for $S>1$, it can be attributed 
to strong quantum fluctuations present at small values of $S$.    \\

{\it   Acknowledgment:}
We thank the
DFG for financial support (grants
DR269/3-1  and RI615/16-1). Computing time at the Leibniz Computing Center
in Garching is gratefully
acknowledged.

\end{document}